\documentclass[aps,twocolumn,nofootinbib,showpacs,prd,aps,10pt]{revtex4}
\usepackage[dvips]{graphicx}
\usepackage[english]{babel}
\usepackage[T1]{fontenc}
\usepackage{mathrsfs}
\usepackage[tbtags]{amsmath}
\usepackage{amssymb}
\usepackage{amsxtra}
\usepackage{amsopn}
\usepackage{latexsym}
\usepackage[mathcal]{eucal}
\usepackage{mathtools}

\newcommand{\BE}{\begin{equation}}
\newcommand{\EE}{\end{equation}}
\newcommand{\BA}{\begin{align}}
\newcommand{\EA}{\end{align}}
\newcommand{\Tr}{\mathrm Tr}
\newcommand{\nn}{\nonumber}

\newcommand{\kkk}{ \frac{{\rm d}^4k}{(2\pi)^4}}

\newcommand{\kkd}{ \frac{{\rm d}^dk}{(2\pi)^d}}

\newcommand{\Rerm}{\mathop{\rm Re}}

\renewcommand{\Im}{\mathop{\rm Im}}

\begin{document}

\title{The gluon propagator in linear covariant $R_\xi$ gauges}

\author{Fabio Siringo and Giorgio Comitini}

\affiliation{Dipartimento di Fisica e Astronomia 
dell'Universit\`a di Catania,\\ 
INFN Sezione di Catania,
Via S.Sofia 64, I-95123 Catania, Italy}

\date{\today}

\begin{abstract}
Explicit analytical expressions are derived for the gluon propagator in a generic linear
covariant $R_\xi$ gauge, by a screened massive expansion for the exact Faddeev-Popov Lagrangian
of pure Yang-Mills theory. At one-loop, if the gauge invariance of the pole structure is 
enforced, the gluon dressing function is entirely and uniquely determined, without any free parameter or external input.
The gluon propagator is found finite in the IR for
any $\xi$, with a slight decrease of its limit value when going from the Landau gauge ($\xi=0$) towards the Feynman gauge
($\xi=1$). An excellent agreement is found with the lattice in the range $0<\xi<0.5$ where the data are
available.
\end{abstract}

\pacs{12.38.Aw, 12.38.Bx, 14.70.Dj, 12.38.Lg}

%12.38.Bx       Perturbative calculations
%12.38.Lg	Other nonperturbative calculations (QCD)
%12.38.Aw	General properties of QCD (dynamics, confinement, etc.)
%14.70.Dj	Gluons
%11.15.Tk       Other nonperturbative techniques (gauge field theories)

%12.38.Gc	Lattice QCD calculations (see also 11.15.Ha Lattice gauge theory)
%11.10.Ef       Field Theory: Lagrangian and Hamiltonian approaches
%11.15.-q	Gauge field theories
%12.20.-m       QED
%11.15.Bt       General properties of perturbation theory (gauge theory)
%12.38.-t	Quantum chromodynamics

\maketitle

\section{Introduction}
Almost all the visible mass in the universe arises from dynamical mass generation, a mechanism that
converts chiral current quarks into constituent quarks, each carrying one third of the proton mass.
The mechanism can be understood as the effect of low-energy gluon clouds dressing the current quark,
so that
the study of the gluon propagator in the IR becomes of paramount importance for a full comprehension of
the mass generation\cite{cornwall,bernard,dono,philip,olive,aguilar04,papa15b}. 
Unfortunately, even in the pure gauge sector, perturbation theory breaks down in the IR and
the results of lattice simulations\cite{olive,cucch07,cucch08,cucch08b,cucch09,bogolubsky,olive09,dudal,binosi12,olive12,burgio15,duarte} 
are regarded as the only benchmark for the continuum approaches\cite{aguilar8,aguilar10,aguilar14,papa15,fischer2009,
huber14,huber15g,huber15b,pawlowski08,pawlowski10,pawlowski10b,pawlowski13,
varqcd,genself,watson10,watson12,rojas,reinhardt04,reinhardt05,reinhardt14}
that have been developed.
Among them, a purely analytical method has been proposed in the last years\cite{ptqcd,ptqcd2,analyt}, 
which is based on a change of the expansion
point of ordinary perturbation theory and provides explicit and very accurate expressions for the gluon propagator in the
Landau gauge\cite{scaling}. 
The method relies on a screened massive expansion, with massive propagators in the internal gluon lines of Feynman graphs,
and is derived from the exact Faddeev-Popov Lagrangian of pure Yang-Mills theory, from first principles, without adding any
phenomenological parameter. The expansion can be seen to emerge from the Gaussian effective potential\cite{journey,varT} which
provides a simple argument for the dynamical mass generation of the gluon and has been also studied at finite temperature\cite{varT,damp}.

In this paper, the massive expansion is extended to the more general case of a linear covariant $R_\xi$ gauge and explicit analytical
expressions are provided for the gluon propagator at any generic value of the gauge-fixing parameter $\xi$, yielding new insight into
the gauge dependence of the propagator, that cannot be extracted by any other method.

Exploring the gauge dependence of the gluon propagator is in itself important in order to individuate the properties that are
gauge invariant and might be directly related to physical observables. 
Despite that, the covariant $R_\xi$ gauge, which is under control
at the perturbative level, is basically unexplored in the IR 
because of convergence problems in lattice calculations\cite{cucch09b,cucch09c,cucch10}. 
Quite recently,
a lattice simulation has been extended up to $\xi=0.5$~\cite{bicudo15}, 
predicting a saturation of the propagator deep in the IR, with very small deviations from the results in the Landau gauge,  
but in strong disagreement with some recent predictions of a continuum study\cite{huber15g}.
On the other hand, the lattice data seem to be in qualitative agreement with the picture emerging by Nielsen identities
in Ref.\cite{papa15}. 
Out of the Euclidean space, no information has been reported so far about the analytic properties in $R_\xi$ gauge.

On general grounds, because of Nielsen identities\cite{nielsen}, 
we know that the poles and the residues of the gluon propagator, i.e. the principal part, 
must be gauge parameter independent\cite{kobes90,breck,dudalR}.
While no information on the existence and properties of the poles can be extracted from lattice calculations in the Euclidean space,
the massive expansion provides explicit analytical expressions that can be continued to the complex plane\cite{analyt}.
Some attempts at reconstructing the spectral functions from the lattice data have been reported\cite{dudal14,cyrol18} and are in qualitative 
agreement with the predictions of the expansion\cite{analyt}. 

At one-loop, by the massive expansion,
a pair of complex
conjugated poles were found in the Landau gauge\cite{damp}, as also predicted by different phenomenological models\cite{dudal08,capri,dudal11}, 
again in strong disagreement with other continuum studies\cite{straussDSE}
based on the truncation of an infinite set of Dyson-Schwinger equations. 
While the genuine nature of the poles was already shown by studying their behavior at finite temperature\cite{damp}, their explicit
gauge invariance would provide further evidence that they are not artifact of the expansion.
Strictly speaking, by changing the expansion point, the Becchi-Rouet-Stora-Tyutin (BRST) symmetry of the quadratic part of the 
Lagrangian is broken in the
expansion and we should not expect that the pole structure might be exactly gauge invariant at any finite order. However, since
the total Lagrangian is not modified, the gauge parameter independence must be recovered if the expansion provides a very good approximation of the
exact propagator. Thus the gauge parameter independence of the pole structure would give a quantitative estimate of the accuracy in the complex plane,
where no comparison with the lattice can be made. 

By the same argument, the massive expansion can be optimized by enforcing the gauge parameter independence of
the whole pole structure, yielding a fully self-contained calculation from first principles, without any adjustable parameter or external input.
Moreover, once optimized in the complex plane, the result is found in excellent agreement with the lattice data in the Euclidean
space, not only in the Landau gauge, but for the whole range, up to $\xi=0.5$, that has been explored in the lattice so far\cite{bicudo15}.
No dramatic difference is found for larger values of $\xi$ and even in the Feynman gauge the gluon propagator is finite in the IR,
with a slight suppression of its saturation value compared to the Landau gauge. Being gauge parameter independent, the principal part of the
propagator might be directly related to physical observables like  glueball masses, 
as recently discussed by a quite general method\cite{dudal11}.
 
The paper is organized as follows:
in Section II the massive expansion of Refs.\cite{ptqcd,ptqcd2} is extended to a generic $R_\xi$ gauge; 
in Section III the expansion is optimized by requiring that the pole structure is gauge parameter independent
as demanded by Nielsen identities; in Section IV the optimized gluon propagator is shown for a wide range of the
gauge parameter $\xi$, including the Feynman gauge ($\xi=1$), and is compared with the available lattice data;
Section V contains a brief discussion of the main results. 
Explicit analytical expressions for the propagator in $R_\xi$ gauge are derived in Appendix A with many details on
the calculation of the graphs.

\section{The massive expansion in $R_\xi$ gauge}

The massive expansion has been first developed in Refs.\cite{ptqcd,ptqcd2} and related to the Gaussian effective potential
in Refs.\cite{journey,varT}. It is based on a change of the expansion
point of ordinary perturbation theory for the exact gauge-fixed Faddeev-Popov Lagrangian of pure Yang-Mills $SU(N)$ theory.
The Lagrangian can be written as
\BE
{\cal L}={\cal L}_{YM}+{\cal L}_{fix}+{\cal L}_{FP}
\label{Ltot}
\EE
where ${\cal L}_{YM}$ is the Yang-Mills term
\BE
{\cal L}_{YM}=-\frac{1}{2} \Tr\left(  \hat F_{\mu\nu}\hat F^{\mu\nu}\right),
\EE
the tensor operator $\hat F_{\mu\nu}$ is 
\BE
\hat F_{\mu\nu}=\partial_\mu \hat A_\nu-\partial_\nu \hat A_\mu
-i g \left[\hat A_\mu, \hat A_\nu\right],
\EE
${\cal L}_{FP}$ is the ghost term
arising from the Faddeev-Popov determinant and
${\cal L}_{fix}$ is the covariant gauge-fixing term 
\BE
{\cal L}_{fix}=-\frac{1}{\xi} \Tr\left[(\partial_\mu \hat A^\mu)(\partial_\nu \hat A^\nu)\right].
\EE
The gauge field operators are
\BE
\hat A^\mu=\sum_{a} \hat X_a A_a^\mu
\EE
where the generators of $SU(N)$ satisfy the algebra
\BE
\left[ \hat X_a, \hat X_b\right]= i f_{abc} \hat X_c,\qquad
f_{abc} f_{dbc}= N\delta_{ad}.
\label{ff}
\EE

In the standard perturbation theory, the total action is splitted as $S_{tot}=S_0+S_I$ where the quadratic part
can be written as
\begin{align}
S_0&=\frac{1}{2}\int A_{a\mu}(x)\delta_{ab} {\Delta_0^{-1}}^{\mu\nu}(x,y) A_{b\nu}(y) {\rm d}^4 x\,{\rm d}^4 y \nn \\
&+\int \omega^\star_a(x) \delta_{ab}{{\cal G}_0^{-1}}(x,y) \omega_b (y) {\rm d}^4 x\, {\rm d}^4 y
\label{S0}
\end{align}
and the interaction is
\BE
S_I=\int{\rm d}^dx \left[ {\cal L}_{gh} + {\cal L}_3 +   {\cal L}_4\right].
\label{SI}
\EE
with the three local interaction terms that read
\begin{align}
{\cal L}_3&=-g  f_{abc} (\partial_\mu A_{a\nu}) A_b^\mu A_c^\nu\nn\\
{\cal L}_4&=-\frac{1}{4}g^2 f_{abc} f_{ade} A_{b\mu} A_{c\nu} A_d^\mu A_e^\nu\nn\\
{\cal L}_{gh}&=-g f_{abc} (\partial_\mu \omega^\star_a)\omega_b A_c^\mu.
\label{Lint}
\end{align}
In Eq.(\ref{S0}), $\Delta_0$ and ${\cal G}_0$ are the standard free-particle propagators for
gluons and ghosts and their Fourier transforms are
\begin{align}
{\Delta_0}^{\mu\nu} (p)&=\Delta_0(p)\left[t^{\mu\nu}(p)  
+\xi \ell^{\mu\nu}(p) \right]\nn\\
\Delta_0(p)&=\frac{1}{-p^2}, \qquad {{\cal G}_0} (p)=\frac{1}{p^2}.
\label{D0}
\end{align}
having used the transverse and longitudinal projectors 
\BE
t_{\mu\nu} (p)=\eta_{\mu\nu}  - \frac{p_\mu p_\nu}{p^2};\quad
\ell_{\mu\nu} (p)=\frac{p_\mu p_\nu}{p^2}
\label{tl}
\EE
where $\eta_{\mu\nu}$ is the metric tensor. 

The massive expansion is obtained by adding a transverse mass term to the quadratic
part of the action and subtracting it again from the interaction, leaving the total action
unchanged.

In some detail, we add and subtract the action term 
\BE
\delta S= \frac{1}{2}\int A_{a\mu}(x)\>\delta_{ab}\> \delta\Gamma^{\mu\nu}(x,y)\>
A_{b\nu}(y) {\rm d}^4\, x{\rm d}^4y
\label{dS1}
\EE
where the vertex function $\delta\Gamma$ is a shift of the inverse propagator
\BE
\delta \Gamma^{\mu\nu}(x,y)=
\left[{\Delta_m^{-1}}^{\mu\nu}(x,y)- {\Delta_0^{-1}}^{\mu\nu}(x,y)\right]
\label{dG}
\EE
and ${\Delta_m}^{\mu\nu}$ is a new massive free-particle propagator 
\begin{align}
{\Delta_m^{-1}}^{\mu\nu} (p)&=
(-p^2+m^2)\,t^{\mu\nu}(p)  
+\frac{-p^2}{\xi}\ell^{\mu\nu}(p).
\label{Deltam}
\end{align}
Adding that term is equivalent to substituting the new massive propagator ${\Delta_m}^{\mu\nu}$ for the 
old massless one ${\Delta_0}^{\mu\nu}$ in the quadratic part.

In order to leave the total action unaffected by the change, we must add the same term in the interaction,
providing a new interaction vertex $\delta\Gamma$.
Dropping all color indices in the diagonal matrices and
inserting Eq.(\ref{D0}) and (\ref{Deltam}) in Eq.(\ref{dG}) the vertex is just the transverse mass shift
of the quadratic part
\BE
\delta \Gamma^{\mu\nu} (p)=m^2 t^{\mu\nu}(p) 
\label{dG2}
\EE
and must be added to the standard set of vertices in Eq.(\ref{Lint}).

The proper gluon polarization $\Pi$ and ghost self energy $\Sigma$ can be evaluated, order by order, by perturbation theory.
In all Feynman graphs the internal gluon lines are replaced by the massive free-particle propagator ${\Delta_m}^{\mu\nu}$ and
all insertions are considered of the (transverse) mass counterterm $\delta \Gamma^{\mu\nu}$ which plays the role of a new two-point vertex. 
It is shown
as a cross in Fig.1 where some two-point self-energy graphs are displayed.
We will refer to the graphs with a cross as {\it crossed} graphs.

Since the total gauge-fixed FP Lagrangian is not modified and because of gauge invariance,
the longitudinal polarization is known exactly and is zero, so that the total polarization
is transverse
\BE
\Pi^{\mu\nu}(p)=\Pi(p)\, t^{\mu\nu}(p)
\label{pol}
\EE
and the (exact) dressed propagators read
\begin{align}
\Delta_{\mu\nu}(p)&=\Delta (p)\,t_{\mu\nu}(p)+\Delta^L (p)\,\ell^{\mu\nu}(p)\nn\\
{\cal G}^{-1}(p)&=p^2-\Sigma (p)
\end{align}
where the transverse and longitudinal parts are
\begin{align}
{\Delta}^{-1} (p)&=-p^2+m^2-\Pi(p)\nn\\
{\Delta^L} (p)&=\frac{\xi}{-p^2}.
\label{DTL}
\end{align}

At tree level, the polarization is just given by the counterterm $\delta \Gamma$ of Eq.(\ref{dG2}),
so that the tree-term $\Pi_{tree}=m^2$ just cancels the mass in the
dressed propagator $\Delta$ of Eq.(\ref{DTL}), giving back the standard free-particle propagator
of Eq.(\ref{D0}).

Finally, summing up the loops and switching to Euclidean space, the transverse dressed propagator
can be written as
\BE
{\Delta}(p)=\left[p^2-\Pi_{loops}(p)\right]^{-1}
\label{dressD}
\EE
where $\Pi_{loops}(p)$ is given by the transverse part of all the loop graphs for the (proper) polarization.

At one-loop, as discussed in Refs.\cite{ptqcd,ptqcd2},
we sum all the graphs with no more than three vertices and no more than one loop, which are displayed in Fig.~1.
In Appendix A, explicit analytical expressions are given for all the polarization graphs of the figure.

\begin{figure}[b] \label{fig:graphs}
\vskip 1cm
\centering
%\hspace*{1cm}
\includegraphics[width=0.25\textwidth,angle=-90]{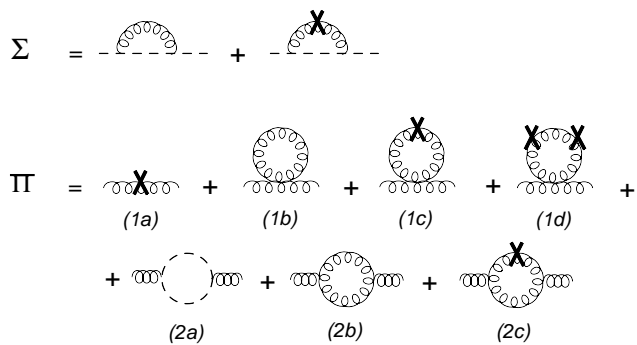}
\caption{Two-point self-energy graphs with no more than three vertices and no more than one loop.} 
\end{figure}

The diverging integrals are made finite by dimensional regularization and can be evaluated in the Euclidean space, by setting $d=4-\epsilon$.
An important feature of the massive expansion is that the crossed graphs cancel all the spurious diverging mass terms exactly, so
that no mass renormalization is required. That is a very welcome feature since there is no bare mass in the original Lagrangian.
At one-loop, as shown in Appendix A, in the $\overline{MS}$ scheme, the diverging part of the proper transverse polarization 
can be written as
\BE
{\Pi}^\epsilon (p)
=\frac{Ng^2}{(4 \pi)^2}\left(\frac{2}{\epsilon}+\log\frac{\mu^2}{m^2}\right)
p^2\left(\frac{13}{6}-\frac{\xi}{2}\right)
\label{PiTeps}
\EE
which is the same identical result of standard perturbation theory\cite{peskin} and ensures that we obtain the correct leading
behavior in the UV where the mass insertions are negligible, as shown in Eq.(\ref{UV}).

As usual the diverging part can be canceled by wave function renormalization, by subtraction
at an arbitrary point. Of course, a finite term $\sim{\rm const.}\times p^2$ arises from the subtraction
and cannot be determined in any way. It also depends on the regularization
scheme and on the arbitrary scale $\mu$, so that its actual value remains somehow arbitrary. 
It basically is the only free parameter of the approximation, as discussed later. 
For an observable particle, the constant would be fixed on 
mass shell, by
requiring that the pole of the propagator is at the physical mass with a residue equal to 1. The confinement of
the gluon has been related to the existence of complex conjugated poles\cite{damp}, so that if, on the one hand,
there is nothing like an observable gluon mass, on the other hand, the analytic properties at the poles and their
gauge parameter independence will be shown to be enough for determining the propagator entirely and uniquely.
 
The finite part of the one-loop proper polarization, as resulting from the sum of all the graphs in Fig.~1,  reads
\BE
\Pi^f(p)=-\frac{3Ng^2}{(4 \pi)^2}\> p^2\> \left[F(s)+\xi\, F_\xi (s) + C\right]
\EE
where $s=p^2/m^2$ is the Euclidean momentum. The functions $F(s)$ and $F_\xi(s)$ are adimensional and do not depend on any
parameter. Their explicit expressions are derived in Appendix A by a detailed calculation of the integrals
and the final result is reported in Eqs.(\ref{Fx}),(\ref{Fxi}).
The constant $C$ arises from the subtraction of the diverging part by wave function renormalization. 
For a generic subtraction point $p=\mu$, the one-loop transverse propagator follows from Eq.(\ref{dressD})
\begin{align}
&\Delta (p)=\nn\\
&\frac{Z_\mu}{p^2+\frac{3Ng^2}{(4 \pi)^2}\> p^2\>\left[F(s)+\xi\, F_\xi (s)-F\left(\frac{\mu^2}{m^2}\right)
-\xi\, F_\xi \left(\frac{\mu^2}{m^2}\right)\right]}
\end{align}
where $Z_\mu$ is the arbitrary finite renormalization constant
$Z_\mu=\mu^2 \Delta (\mu)$. Finally, the propagator can be written as
\BE
\Delta (p)=\frac{Z}{p^2\left[F(s)+\xi\, F_\xi (s) + F_0\right]}
\label{prop1}
\EE
where the coupling and all other constants are absorbed by a finite renormalization factor $Z$ and
the new constant $F_0$ which depend on the subtraction point $\mu$ according to
\begin{align}
Z&=\frac{(4 \pi)^2 Z_\mu}{3Ng^2}\nn\\
F_0&=\frac{(4 \pi)^2 }{3Ng^2}-F\left(\mu^2/m^2\right)-\xi\, F_\xi \left(\mu^2/m^2\right).
\label{F0def}
\end{align}

Eq.(\ref{prop1}) provides an explicit analytical expression for the one-loop gluon propagator. It contains 
three parameters: $m$, $Z$ and $F_0$.
However, the finite renormalization factor $Z$ is irrelevant, while $m$ is the unique energy scale.
Since the exact Lagrangian does not contain any energy scale, $m$ cannot be determined by the theory:
the mass parameter $m$ determines the overall energy scale and can only be fixed by comparison with some physical observable.
That is not a limitation of the approximation but is a standard feature of Yang-Mills theory.
Moreover, being just a scale parameter, the mass $m$ is not a physical or dynamical mass and is not even required to be gauge invariant.
We will use the energy scale of the lattice and fix $m$ by comparison with the data of simulations in the Landau gauge.
Thus, the only free parameter in Eq.(\ref{prop1}) is the constant $F_0$ which is related to the arbitrary ratio $\mu/m$.
Since the result does depend on $F_0$, the expansion must be optimized by a criterion for determining
the best $F_0$, yielding a special case of optimized perturbation theory by variation of the renormalization scheme, a method
that has been proven to be very effective for the convergence of the expansion\cite{stevensonRS}.

Assuming that the expansion converges more quickly for an optimal value of $F_0$, the one-loop result might be very
close to the exact result for a special choice of the constant. That is shown to be the case in Refs.\cite{ptqcd,ptqcd2,analyt,scaling}
where an excellent agreement with the lattice is found in the Landau gauge.
Unfortunately, the available data are not fully consistent and a best fit yields slightly different values of $F_0$ and $m$
for different data sets, as shown in Table I. 
The deviations might be related to a slightly different choice of units as recently discussed in Ref.\cite{boucaud}.
We can extract a global average $F_0\approx -0.9\pm 0.1$.
Of course, the actual value of the constant $F_0$ depends on
the details of the definition of the functions $F(s)$, $F_\xi(s)$ which are evaluated up to
an (omitted) arbitrary additive constant in Appendix A. In this paper, all the values of $F_0$  refer to the
definition given by Eqs.(\ref{Fx}),(\ref{Fxi}) for those functions.

\begin{table}[ht]
\centering % used for centering table
\begin{tabular}{c c c c c} % centered columns (4 columns)
\hline\hline %inserts double horizontal lines
Data set & $N$ & $F_0$ & $m$ (GeV) & $Z$ \\ [0.5ex] % inserts table
%heading
\hline % inserts single horizontal line
Duarte et al.\cite{duarte}              & $SU(3)$ & -0.887   &  0.654  & 2.631 
\\ % inserting body of the table
Bogolubsky et al.\cite{bogolubsky}      & $SU(3)$ & -1.035   &  0.733  & 3.360  
\\ [1ex] % [1ex] adds vertical space
\hline %inserts single line
Cucchieri,Mendes\cite{cucch08,cucch08b} & $SU(2)$ & -0.743   &  0.859  & 1.737   
\\ [1ex] % [1ex] adds vertical space
\hline %inserts single line
\end{tabular}
\label{tableI} % is used to refer this table in the text
\caption{Parameters of Eq.(\ref{prop1}) optimized by the $SU(3)$ data of Ref.\cite{duarte} (in the range 0 - 4 GeV) and
Ref.\cite{bogolubsky} (0 - 2 GeV), and by the $SU(2)$ data of Refs.\cite{cucch08,cucch08b} (0 - 2 GeV).}
\end{table}

\begin{figure}[b] \label{fig:Delta}
\centering
\includegraphics[width=0.35\textwidth,angle=-90]{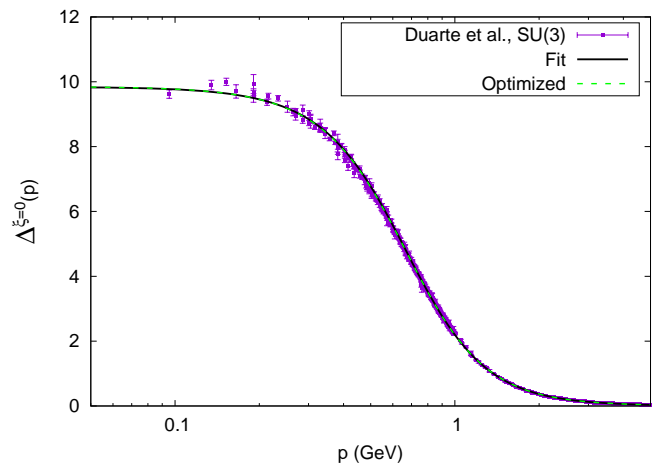}
\caption{The one-loop transverse gluon propagator $\Delta (p)$ of Eq.(\ref{prop1}) is shown for the best fit parameters $F_0=-0.887$, $m=0.654$ GeV
in the Landau gauge $\xi=0$ (solid line), together with the lattice data of Ref.\cite{duarte}.
The broken line is the same propagator obtained by Eq.(\ref{prop1}) with the optimized parameters $F_0=-0.876$, $m=0.656$ GeV determined by
the gauge parameter independence of the pole structure in Section III.}
\end{figure}

In the Landau gauge, the best agreement is found for the data set of Ref.\cite{duarte}, with a best fit parameter $F_0=-0.887$
and a mass scale $m=0.654$ GeV, evaluated in the range $0<p<4$ GeV. The resulting gluon propagator is shown in Fig.~2 together 
with the lattice data.

\section{Optimization from first principles}

If the expansion is optimized in the Euclidean space, by a direct comparison with the lattice data, 
any control of the approximation is lost in Minkowski space and one 
might wonder how robust the optimal choice would be when continued to the complex plane. 
Moreover, a self-contained optimization strategy, which does not require any external input, would 
be essential for exploring new aspects that are out of the reach of lattice calculations.
In this section, we show that the expansion can be optimized from first principles in the complex plane 
by enforcing some
general exact analytic properties that arise from the BRST invariance of the gauge-fixed Lagrangian.

The Nielsen identities\cite{nielsen} are exact equations connecting the gauge parameter dependence of
some correlation functions with other Green functions. Their proof follows from the BRST invariance of
the Faddeev-Popov Lagrangian, Eq.(\ref{Ltot}), which has not been modified by our change of the expansion point.
They have been used as a tool for establishing general invariance properties of the pole structure 
in QCD\cite{breck} and in other Yang-Mills theories\cite{capri}.

Following the detailed derivation of Ref.\cite{breck},
the exact transverse projection of the gluon propagator $\Delta (p)$ must satisfy the Nielsen identity

\BE
\frac{\partial}{\partial \xi} \frac{1}{\Delta(p)}= G^T(p)\,\left[\frac{1}{\Delta(p)}\right]^2 
\label{nielsen}
\EE
where, omitting the diagonal color indices, $G^T(p)$ is the transverse component
\BE
G^T(p)=\frac{t_{\mu\nu}(p)}{3} G^{\mu\nu}_{aa}(-p,p,0)
\label{GT}
\EE
of the Green function $G^{\mu\nu}_{ab}(-p,p,0)$ which is defined as
\begin{align}
&G^{\mu\nu}_{ab}(-p,p,0)=\int{\rm d}^4x\,{\rm d}^4y\> e^{ip\cdot(x-y)}\,\times\nn\\
&\qquad\qquad\times \langle 0\vert T\left[D^\mu\omega_a(y) A^\nu_b(x) \omega^\star_c(0)B_c(0) \right]\vert 0\rangle
\end{align}
in terms of the Nakanishi-Lautrup auxiliary field $B_a$ and of the covariant derivative of the ghost field
$D^\mu \omega_a$.
If the gluon propagator has a pole in the complex plane at $p^2=p^2_0(\xi)$, then the inverse propagator has a zero
and we can write the identities
\BE
\frac{1}{\Delta\left(p_0(\xi)\right)}=0;\quad  \frac{\rm d}{{\rm d}\xi}\frac{1}{\Delta\left(p_0(\xi)\right)}=0.
\label{pole}
\EE
Then, the vanishing of the right hand side of Eq.(\ref{nielsen}) at $p=p_0(\xi)$ says that the
partial derivative is also zero and
the pole $p_0$ must be gauge parameter independent
\BE
\frac{\rm d}{{\rm d}\xi} p_0(\xi)=0.
\label{dp0}
\EE

By the same argument, the residues at the poles are also gauge parameter independent\cite{dudalR}.
In fact, if we differentiate Eq.(\ref{nielsen}) with respect to $p^2$
\begin{align}
\frac{\partial}{\partial \xi}\left[ \frac{\rm d}{{\rm d} p^2} \frac{1}{\Delta(p)}\right]&= 
\left[\frac{\rm d}{{\rm d} p^2}\,G^T(p)\right]\,\left[\frac{1}{\Delta(p)}\right]^2 \nn\\
&+2G^T(p)\,\frac{1}{\Delta(p)}
\left[ \frac{\rm d}{{\rm d} p^2} \frac{1}{\Delta(p)}\right],
\label{nielsenR}
\end{align}
the right hand side vanishes at $p=p_0$ because of Eq.(\ref{pole}), so that the residue $R$,
defined as
\BE
R=\lim_{p\to p_0}\, \Delta(p) (p^2-p_0^2)=\lim_{p\to p_0} \left[ \frac{\rm d}{{\rm d} p^2} \frac{1}{\Delta(p)}\right]^{-1},
\label{Rdef}
\EE
satisfies the exact equation
\BE
\frac{\partial}{\partial \xi} R=0.
\label{dR0}
\EE
We conclude that, for the gauge-fixed Yang-Mills Lagrangian, the principal part $\Delta^P$ of the exact gluon propagator 
\BE
\Delta^P(p)=\frac{R}{p^2-p_0^2}+\frac{R^\star}{p^2-{p_0^\star}^2}
\label{princ}
\EE
must be gauge parameter independent.
The argument fails if $G^T(p)$ has a pole in $p=p_0$, which is usually not the case. 

In the quadratic part of the Lagrangian, the BRST symmetry is broken by the mass term that has been added and has been subtracted
again from the interaction. Thus, while the total Lagrangian is BRST invariant, the symmetry is broken at any finite order of
the massive expansion. For that reason, we do not expect that the one-loop propagator might satisfy the Nielsen identity exactly.
However, the closer we reach to the exact result, the better is expected to be the agreement with the exact identities.
Thus, we can exploit the dependence on the parameters $F_0$, $m$ in Eq.(\ref{prop1}) and optimize the expansion by requiring
that the pole structure of the propagator is gauge parameter independent. That is equivalent to an optimal choice of the
subtraction point $\mu/m$, which is usually fixed on mass shell for an observable particle. Without any observable gluon mass
at hand, the invariance of the poles and residues turns out to be enough for determining the one-loop gluon propagator entirely and for
any choice of the gauge parameter.

For a generic choice of the gauge parameter $\xi$, the optimal parameters can
be regarded as functions $F_0(\xi)$, $m(\xi)$,  to be determined by the requirement that the pole and the residue do not depend
on $\xi$. Of course, the finite renormalization factor $Z$ remains arbitrary and has no physical relevance.
Let us denote by $\Psi(z,\xi,F_0,m)$ the inverse dressing function in Eq.(\ref{prop1})
\BE
\Psi(z,\xi,F_0,m)=F(-z^2/m^2)+\xi\, F_\xi (-z^2/m^2) + F_0
\label{Psi}
\EE
which is an analytic function of the complex variable $z=x+iy$. On the imaginary axis, for $x=0$, 
$p_E^2=-z^2=y^2$ is the Euclidean momentum. On the real axis, for $y=0$, we recover the Minkowskian momentum
$p_M^2=z^2=x^2$. Thus, the variable $z$ is the analytic continuation of the physical momentum $p_M$.
The pole $z_0^2=-p_0^2$ is a zero of the inverse dressing function $\Psi$ and must satisfy the equation $\Psi(z_0,\xi,F_0,m)=0$.
The gauge parameter independence of the pole requires that
\BE
\Psi\left(z_0,\xi_1, F_0(\xi_1), m(\xi_1)\right)=
\Psi\left(z_0,\xi_2, F_0(\xi_2), m(\xi_2)\right)
\EE
yielding a set of two coupled real equations for the real and imaginary parts.
The equations can be solved for $F_0(\xi_2)$ and $m(\xi_2)$ from a given
initial value $F_0(\xi_1)$, $m(\xi_1)$.
Taking the Landau gauge as the initial point $\xi_1=0$ and
fixing a scale $m_0=m(0)$ as energy units, the functions $F_0(\xi)$ and $m(\xi)$ are determined
for any value of the gauge parameter $\xi$ from the initial value $F_0(0)$ which remains the only free 
parameter. Thus, we can encode the gauge parameter independence of the pole in the optimized propagator
and evaluate it for any value of the parameter $\xi$.
The functions $F_0(\xi)$, $m^2(\xi)$ are shown in Fig.~3 and Fig.~4 for different choices of the initial value $F_0(0)$ in the Landau gauge.

In the range $-2<F_0(0)<0$, the gluon propagator of Eq.(\ref{prop1}) has a single pair of complex conjugated poles, 
while other values of $F_0(0)$, out of that range, seem to be unphysical.
For $F_0(0)<-2$ the expression in Eq.(\ref{prop1}) has poles in the Euclidean space and changes sign at the poles, on the positive $s$ axis.
Moreover, according to Eq.(\ref{F0def}), the coupling $g^2$ would become negative in that range because the minimal value of $F(s)$ is 
$\approx 2$.
For $F_0(0)>0$  the coupling $g^2$ becomes very small in Eq.(\ref{F0def}) for any $\mu$ and the pole topology becomes very different. 
As discussed in the previous section, in the Landau gauge, the best agreement with the lattice is found for $F_0\approx -0.9$ which is 
at the center of the physical allowed range.

\begin{figure}[t] \label{fig:mass}
\centering
\includegraphics[width=0.35\textwidth,angle=-90]{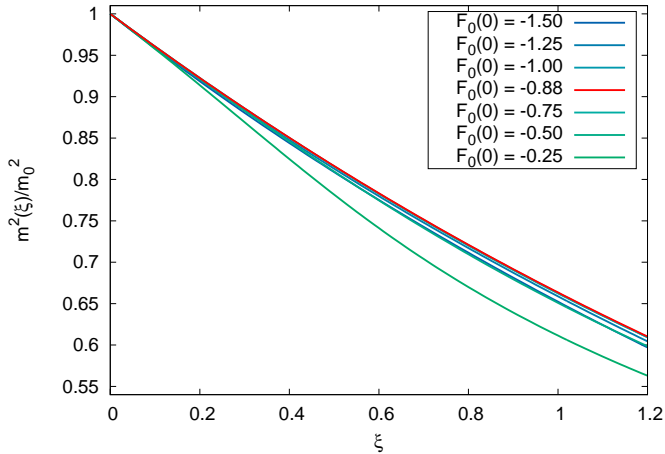}
\caption{The mass parameter ratio $m^2(\xi)/m_0^2$ as a function of the gauge parameter $\xi$ for different initial values of $F_0(0)$.
The red line is obtained for the optimal value $F_0(0)=-0.876$.}
\end{figure}

\begin{figure}[t] \label{fig:F0}
\centering
\includegraphics[width=0.35\textwidth,angle=-90]{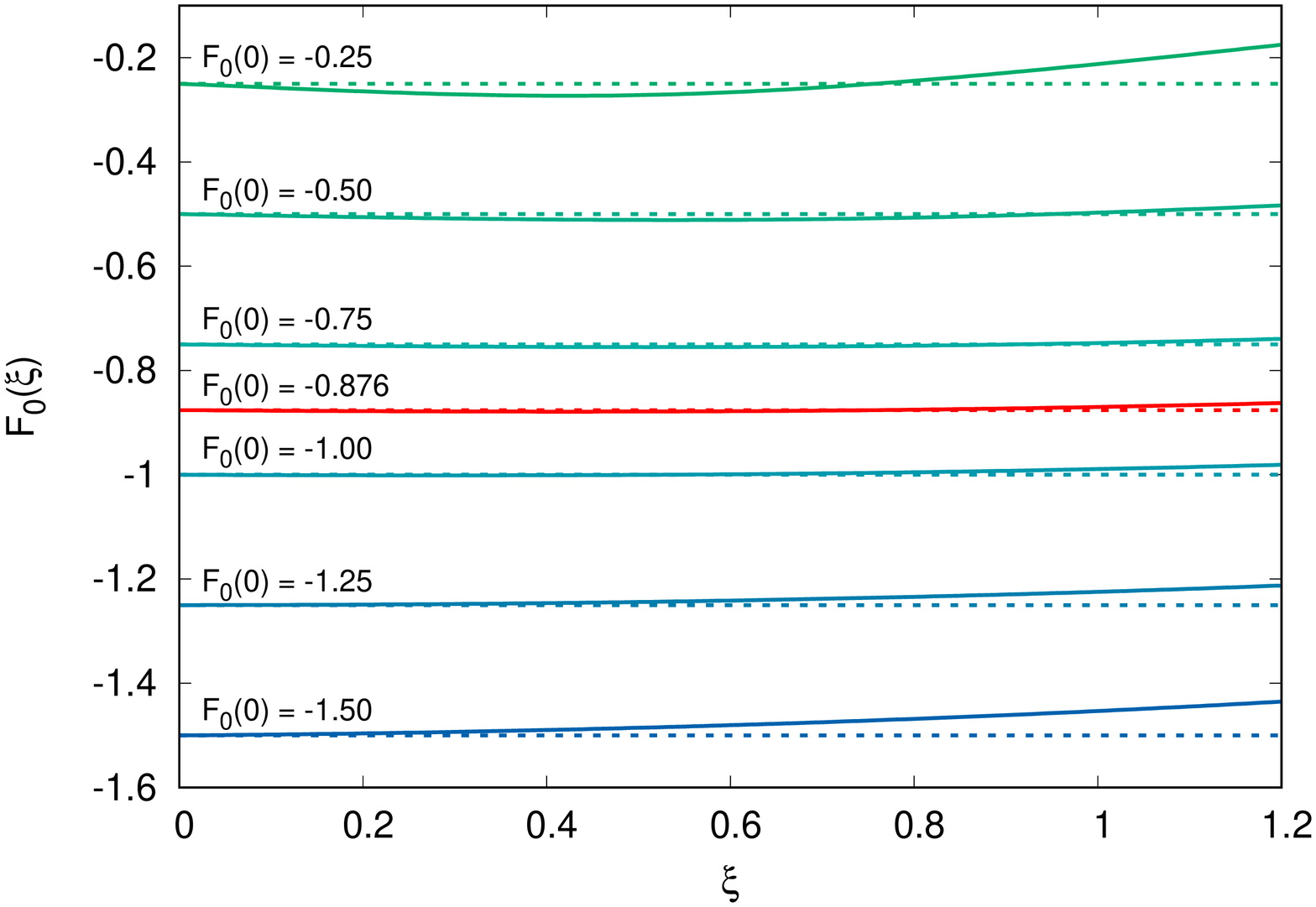}
\caption{The parameter $F_0(\xi)$ as a function of the gauge parameter $\xi$ for different initial values of $F_0(0)$.
The red line is obtained for the optimal value $F_0(0)=-0.876$.}
\end{figure}

\begin{figure}[b] \label{fig:contour}
\centering
\includegraphics[width=0.35\textwidth,angle=-90]{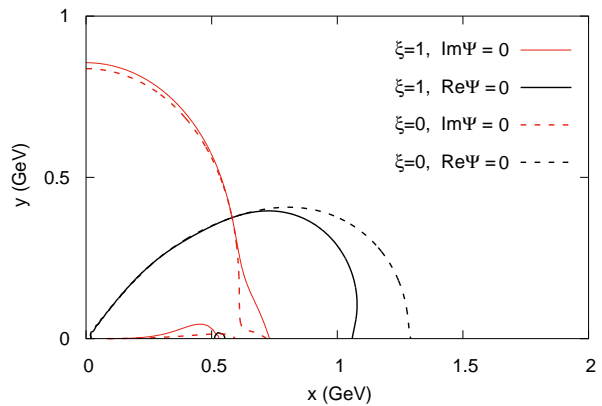}
\caption{Contour plots of $\Rerm\Psi=0$, $\Im\Psi=0$ in the complex plane $z=x+iy$ for $\xi=1$ (solid lines) and $\xi=0$ (dashed lines), 
with $F_0(0)=-0.876$ and $m_0=0.656$ GeV (see Table II).
The curves are approximately tangent (i.e. $\theta\approx 0$) at the intersection point $z_0$ (the pole) whenever $F_0\approx-0.9$.} 
\end{figure}

\begin{figure}[b] \label{fig:phase}
\vskip 1cm
\centering
\includegraphics[width=0.45\textwidth,angle=0]{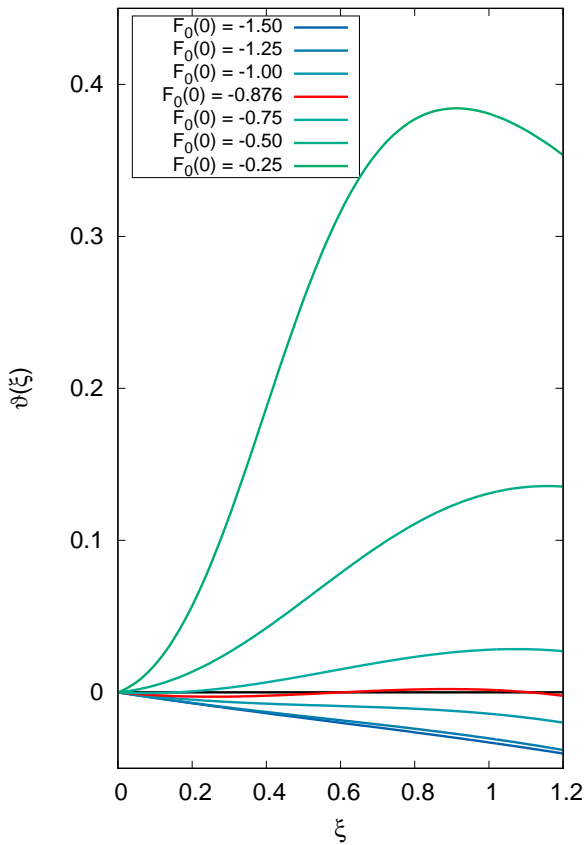}
\caption{The phase change $\theta$ of the residue is shown as a function of $\xi$ for different initial values of $F_0(0)$.
The red line is obtained for the optimal value $F_0(0)=-0.876$.}
\end{figure}

It is remarkable that, close to the best fit value $F_0(0)\approx -0.9$, the contour lines
$\Rerm\Psi=0$, $\Im\Psi=0$,  at the crossing point $z_0$ (the pole), are basically not rotated by any change
of $\xi$. That can be seen in Fig.~5 where the contour lines are displayed for $\xi=0$
and $\xi=1$ and are shown to be approximately tangent at the intersection point.
In other words, when the initial value $F_0(0)$ approaches
the best fit value $F_0(0)\approx -0.9$, the conformal map $z_1\to z_2$, defined by
\BE
\Psi\left(z_1,\xi_1, F_0(\xi_1), m(\xi_1)\right)=\Psi\left(z_2,\xi_2, F_0(\xi_2), m(\xi_2)\right)
\EE
becomes a local identity at the fixed point (the pole $z_0^2=-p_0^2$).  
Denoting by $\theta$ the rotation angle of the contour lines in the map and
setting $\xi_1=0$, $\xi_2=\xi$, we can write
\BE
\theta(\xi)={\rm Arg} \left\{\frac
{\displaystyle{  \frac{\rm d}{{\rm d} z} } \Psi\left(z,0, F_0(0), m(0)\right) } 
{\displaystyle{ \frac{\rm d}{{\rm d} z} } \Psi\left(z,\xi, F_0(\xi), m(\xi)\right) }
\right\}_{\displaystyle{z=z_0}}
\EE
and because of Eq.(\ref{Rdef}), the angle $\theta$ gives the phase change of the residue $R$ which
can be written, as a function of $\xi$,
\BE
R(\xi)=R(0)\, e^ {i\theta(\xi)}
\EE
since the modulus $\vert R\vert$ can  always be made invariant by an appropriate choice of the
real renormalization constant $Z(\xi)$. Explicit analytical expressions for the derivative of $\Psi$ are
reported in Eqs.(\ref{der1}),(\ref{der2}) of Appendix A. 

We observe that the angle $\theta$ is not  exactly zero, so
that in general, the Nielsen identity Eq.(\ref{nielsen}) and its consequences Eqs.(\ref{dp0}),(\ref{dR0}) cannot be
all satisfied. 
However, as shown in Fig.~5 and Fig.~6, the angle $\theta$ becomes very small, for a wide range of $\xi$, if the initial
constant $F_0(0)$ is close to the value $F_0\approx -0.9$ which already described the lattice data
very well in the Euclidean space. In other words, the optimal propagator in the Euclidean space is also the one that
best satisfies the Nielsen identity in the complex plane, giving us confidence in the general accuracy of the approximation.
We must mention that averaging over Gribov copies might break BRST invariance in the lattice. However, we are assuming that
the Nielsen identities are not seriously affected in lattice calculations.

Reversing the argument, the expansion can be optimized in a self-contained way, by first principles and without any external
input, by assuming that the best choice for the initial constant $F_0(0)$ is the one that makes the angle $\theta$ smaller in
a wider range of $\xi$. Even if there are no technical reasons for limiting the value of the gauge parameter, we expect that
perturbation theory would be more effective when $\xi$ is small and the expansion might be out of control for very large $\xi\gg 1$.
Prudentially, the present study is limited to the range $\xi<1.2$, including the Feynman gauge.

The minimal phase deviation is observed for the initial value $F_0(0)=-0.876$. As shown in Fig.~6,
for that choice, the phase $\theta$ fluctuates around zero in the whole range $0<\xi<1.2$, with very small deviations which
are less than $0.003$. Nevertheless, no fine tuning is required since $\theta$ is very small around $F_0(0)\approx -0.9$ and any 
slight change of $F_0(0)$ can be compensated by an appropriate choice of $Z$ and $m$. 
In fact, as shown in Fig.~2,  when the present new set of first-principle 
optimal parameters are inserted in Eq.(\ref{prop1}),  the propagator is indistinguishable from the previous one that was obtained by
a best fit of the lattice data.
Actually, the optimal initial value $F_0(0)$ not only minimizes the phase deviation $\theta(\xi)$, but also makes $m^2(\xi)$ stationary
and maximal for any fixed $\xi$, as shown in Fig.~3 where the optimal curve is plotted as a red line. That is a geometric consequence
of the pole being the tangency point in Fig.~5. Moreover, the optimal function $F_0(\xi)$ is the most gauge parameter invariant curve
in  Fig.~4 (shown as a red line).

The optimal parameters are summarized in Table II together with very accurate polynomial interpolation formula for the optimal functions
$F_0(\xi)$, $m^2(\xi)$. Extracting the energy scale $m_0=m(0)=0.656$ GeV from the lattice data of Ref.\cite{duarte} in the Landau gauge, 
the invariant pole is found at $x_0=M=0.581$~GeV and $y_0=\gamma=0.375$~GeV, which might be regarded as the physical mass and the damping rate of
the quasigluon, respectively, as discussed in Ref.\cite{damp}.

\begin{table}[ht]
\centering % used for centering table
\begin{tabular}{c} % centered columns (1 columns)
\hline\hline %inserts double horizontal lines
OPTIMIZATION BY GAUGE INVARIANCE
\\ [1ex] % [1ex] adds vertical space
\hline
$F_0(0)=-0.876$,\quad $m_0=m(0)=0.656$~GeV,\quad $Z(0)=2.684$
\\ [1ex] % [1ex] adds vertical space
$\vert \theta(\xi)\vert < 2.76\cdot 10^{-3}$, \qquad $0<\xi<1.2$
\\ [1ex] % [1ex] adds vertical space
\hline
$F_0(\xi)\approx -0.8759-0.01260\xi+0.009536\xi^2+0.009012\xi^3$
\\ [1ex] % [1ex] adds vertical space
$m^2(\xi)/m_0^2\approx 1-0.39997\xi+0.064141\xi^2$
\\
\hline
$z_0/m_0=0.8857+0.5718\, i$, \quad $t_R=\Im R(0)/\Rerm R(0)=3.132$
\\ [1ex] % [1ex] adds vertical space
$M=0.581$~GeV, $\gamma=0.375$~GeV\quad (invariant pole) 
\\
\hline
\end{tabular}
\label{tableII} % is used to refer this table in the text
\caption{Set of optimal parameters, obtained by enforcing the gauge parameter independence of the pole structure in the range
$0<\xi<1.2$. The energy
scale $m_0$ and the finite renormalization constant $Z(0)$ are determined by the data of Ref.\cite{duarte} which are shown
in Fig.~2.}
\end{table}

\section{The propagator at $\xi\not=0$}

In the Euclidean space, the gluon propagator can be evaluated analytically by Eq.(\ref{prop1}), for any value of the gauge parameter
$\xi$, inserting the optimal parameters of Table II which enforce the gauge parameter independence of the pole structure in
the complex plane. In order to compare with the available lattice data of Ref.\cite{bicudo15}, the finite renormalization constant $Z$
is fixed by the same momentum subtraction scheme of that work, i.e. requiring that $\mu^2\Delta(\mu)=1$ for any $\xi$ and
taking the same renormalization point $\mu=4.317$ GeV.
That is equivalent to taking the constant $Z$ in Eq.(\ref{prop1}) to be
$Z=\Psi(i\mu,\xi,F_0,m)$.

\begin{figure}[t] \label{fig:prop}
\centering
\includegraphics[width=0.35\textwidth,angle=-90]{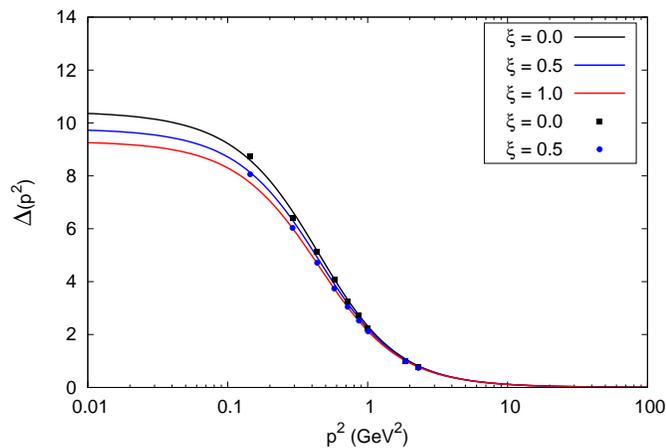}
\caption{The gluon propagator of Eq.(\ref{prop1}) is evaluated as a function of the Euclidean momentum $p^2$ 
with the first-principle optimized parameters of Table II,
for $\xi=0$, $0.5$, $1$ and renormalized at $\mu=4.317$ GeV. The points are the lattice data of Ref.\cite{bicudo15}.}
\end{figure}

\begin{figure}[t] \label{fig:propratio}
\centering
\includegraphics[width=0.35\textwidth,angle=-90]{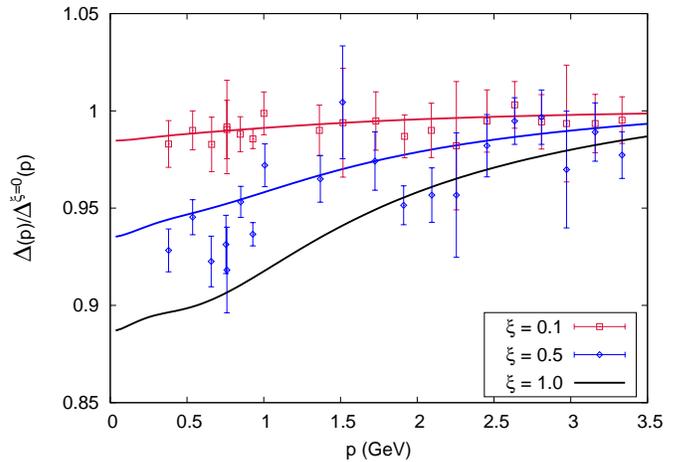}
\caption{The ratio $\Delta(p)/\Delta^{\xi=0}(p)$ as a function of the Euclidean momentum $p$ with the 
first-principle optimized parameters of Table II, for $\xi=0.1$, $0.5$ and $1$.
The bars are the lattice data of Ref.\cite{bicudo15}.}
\end{figure}

\begin{figure}[b] \label{fig:dressing}
\centering
\includegraphics[width=0.35\textwidth,angle=-90]{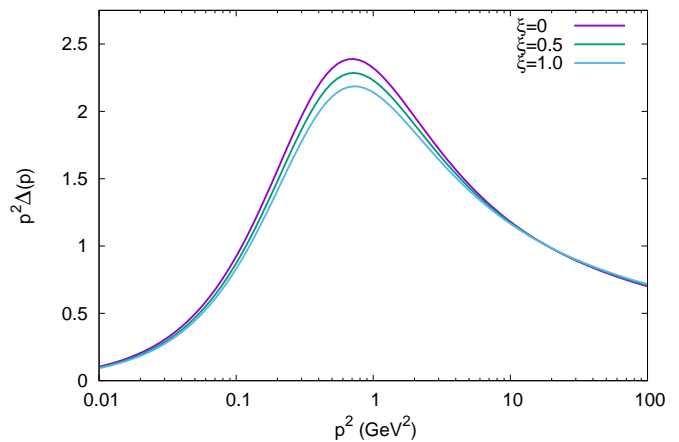}
\caption{The dressing function $p^2 \Delta(p)$ as a function of the Euclidean momentum $p^2$ for the same
parameters of Fig.~7.}
\end{figure}

The gluon propagator is shown in Fig.~7, for several values of the gauge parameter $\xi$, together with some data points extracted from
Ref.\cite{bicudo15}.The agreement with the data is very good in the limited range $\xi<0.5$ where they are available. For $\xi\not=0$,
the propagator is slightly suppressed in the IR compared with the Landau gauge. We must mention that previous continuum studies,
based on the truncation of an infinite set of exact Dyson-Schwinger equations, reached contrasting and ambiguous results. While a
strong dependence on the gauge parameter was predicted in Ref.\cite{huber15g}, with large deviations from the Landau gauge, a qualitative
agreement with the lattice was reported in Ref.\cite{papa15} by the aid of exact Nielsen identities which seem to play a key role. 
The gauge dependence was found small but no quantitative prediction could be made and even the sign of the change was not defined by that method. 

As shown in Fig.~7, up to and beyond the Feynman gauge ($\xi=1$), no dramatic change occurs and the suppression of the propagator increases very
smoothly with the increasing of $\xi$. The change can best be seen by evaluating the ratio between $\Delta (p)$ at $\xi\not=0$ and
at $\xi=0$, as shown in Fig.~8 together with the lattice data of Ref.\cite{bicudo15}. Even if the lattice calculation is plagued
by large statistical errors, with scattered data and large error bars, the optimized propagator seems to be in quantitative agreement
with the data and reproduces the correct trend predicted by the lattice. We stress that the curves are not a fit of the data
and the agreement is reached from first principles without any adjustable parameter.

The dressing function is shown in Fig.~9. As predicted by the lattice\cite{bicudo15}, 
the maximum is basically fixed at the same energy for any $\xi$.
We argue that the Nielsen identity gives the correct scale factor $m(\xi)/m(0)$ that keeps the maximum fixed, at variance 
and in strong contrast with the continuum calculation of Ref.\cite{huber15g} which might miss that important constraint.

In the studied range of $\xi$, the whole principal part of the propagator in Eq.(\ref{princ}) is basically invariant up
to a finite renormalization factor. The pole $p_0$ is fixed at the value of Table II, while the 
phase of the residue is ${\rm Arg}\,R(\xi)=1.262+\theta(\xi)$ where $\vert \theta(\xi)\vert < 2.75\cdot 10^{-3}$, yielding
the ratio $t_R=\Im R(\xi)/\Rerm R(\xi)=3.132\pm 0.03$. This ratio is important for determining the explicit parameters of the rational
part Eq.(\ref{princ}) which has been derived at tree level by other phenomenological models 
like the refined Gribov-Zwanziger model\cite{dudal08,capri,dudal11}. Being gauge parameter independent, the parameters of the rational part 
might be directly related to physical observables or condensates\cite{dudal10,dudal18} and a recent general method has been proposed 
for extracting information on the glueball masses\cite{dudal11}.
Using the notation of Ref.\cite{dudal18}, the principal part of the propagator, Eq.(\ref{princ}), can be written as
\BE
\Delta^P(p)=Z_{GZ}\,\frac{p^2+M_1^2}{p^4+M_2^2p^2+M_3^4}
\label{princ2}
\EE
where
\begin{align}
Z_{GZ}&=2\Rerm R\nn\\
M_1^2&=M^2-\gamma^2+2M\gamma\,t_R=1.562\>{\rm GeV}^2 \nn\\
M_2^2&=2(M^2-\gamma^2)=0.394\>{\rm GeV}^2 \nn\\
M_3^4&=(M^2+\gamma^2)^2=0.229\>{\rm GeV}^4
\end{align}
having made use of the optimized parameters of Table II. Below 1 GeV, the masses $M_i$ seem to be compatible with the statistical analysis of 
Ref.\cite{dudal18}, even if the simple rational part $\Delta^P$ was used in that work for a fit of the lattice data, ignoring
the corrections which are included in the present optimized one-loop propagator. In fact, the corrections are gauge dependent and very
small below 1 GeV, as already shown in the Landau gauge by a direct evaluation of the spectral function\cite{analyt,dispersion}.

The Schwinger function $\Delta (t)$ can be evaluated by a numerical integration, as a function of the
Euclidean time $t$, according to its definition
\BE
\Delta(t)=\int_{-\infty}^{+\infty} \frac{{\rm d} p_4}{2\pi}\, e^{ip_4t}\,
\Delta(\vec p=0, p_4)
\label{SF}
\EE
and is shown in Fig.~10 for different values of the gauge parameter. In the Landau gauge, the Schwinger function
is found in qualitative agreement with the result of Ref.\cite{alkofer}, with a positivity violation that
occurs above the point $t=t_0\approx 5.8$ GeV$^{-1}$ where the function crosses the zero and becomes negative.
The scale $t_0$ is roughly the size of a hadron and in Ref.\cite{alkofer} it was conjuctered to be a physical 
gauge-invariant scale at which gluon screening occurs. Actually, as shown in Fig.~10, the crossing point $t_0$ is found
to be almost gauge parameter independent. Moreover, the large $t$ behavior seems to be dominated by the singularities and the whole Schwinger function is very well approximated 
by inserting in Eq.(\ref{SF}) the simple principal part $\Delta^P(p)$ of Eq.(\ref{princ}), which is
gauge parameter independent, yielding the analytical result
\BE
\Delta^P(t)=\left[\frac{\vert R\vert}{\sqrt{M^2+\gamma^2}}\right] e^{-Mt}
\cos\left(\gamma t-\theta+\arctan\frac{\gamma}{M}\right)
\label{SFP}
\EE
which is shown in Fig.~10 as a broken line.

\begin{figure}[t] \label{fig:schwinger}
\centering
\includegraphics[width=0.35\textwidth,angle=-90]{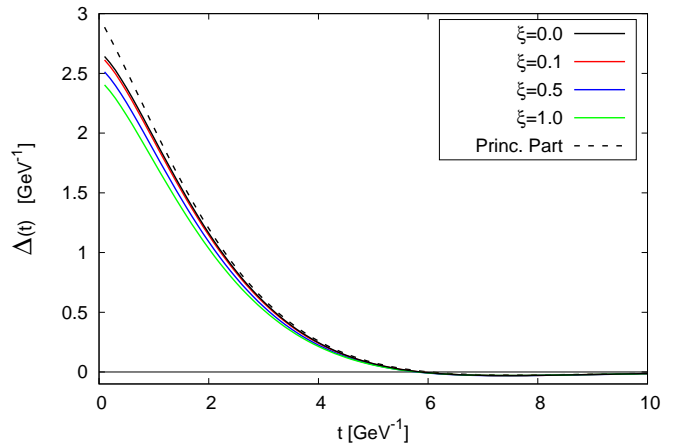}
\caption{The Schwinger function is shown, as a function of the Euclidean time $t$, for different values of the gauge parameter. The broken line is the analytical result of Eq.(\ref{SFP}) which is obtained by the 
principal part of the gluon propagator.}
\end{figure}

We cannot end this section without a brief discussion of the spectral function, which has attracted great interest\cite{dudal14,cyrol18}
even if its physical content is quite unclear in presence of complex poles and confinement. 
In fact, the usual  K\"allen-Lehmann representation must be replaced by the more general integral representation\cite{dispersion}
\begin{align}
\Rerm \Delta (p^{\,2})&= \Delta^P(p^{\,2})+{\rm P.V.}\int_0^{+\infty}\frac{\rho(\mu^2)}{p^2-\mu^2}\> {\rm d} \mu^2\nn\\
\rho(p^2)&=-\frac{1}{\pi} \Im \Delta(p^2+i\epsilon)
\label{KL}
\end{align}
where the spectral function $\rho(p^2)$ is gauge dependent and does not contain any information on the gauge parameter independent principal part
$\Delta^P$ which must be added to the integral for reproducing the whole propagator. Moreover, $\rho(p^2)$ is even not
positive defined for a confined particle. In the Landau gauge, the spectral function was evaluated by the massive expansion in
Ref.\cite{analyt} and the dispersion relation of Eq.(\ref{KL}) was checked in Ref.\cite{dispersion} by a numerical integration.
The integral provides the difference between the principal part and the whole propagator, so that the difference can be large only if
the total weight which comes from the integration of $\rho(p^2)$ is large. Moreover, $\rho(p^2)$ changes sign and the contributions
arising from different signs can partially cancel.

The one-loop spectral density can be easily evaluated by the explicit expression of the propagator, Eq.(\ref{prop1}), using the
optimal parameters of Table II, and is shown in Fig.~11 for different values of the gauge parameter $\xi$. It has some gauge
dependent features, like a cusp at the two-particle threshold $p=2m(\xi)$ and a finite spike at $p\approx m(\xi)$. In the Landau gauge, 
the spike is just a smooth maximum but is enhanced for $\xi>0.08$ by the appearance of a gauge dependent pole near 
the real axis, at $x\approx m(\xi)$. Some details of the finite peak on the real axis are shown in Fig.~12. 
Apart from the peak, the spectral density is very small
and even the peak area gives a small contribution to the integral in Eq.(\ref{KL}) 
because of the change of sign that occurs just at the peak,
in agreement with a confinement scenario. While the peak resembles the spike which was observed in Ref.\cite{straussDSE}, its physical nature
is unclear and is certainly related to the nature of the new gauge dependent pole which might be an artifact of the one-loop approximation.  

\begin{figure}[t] \label{fig:spectral}
\centering
\includegraphics[width=0.35\textwidth,angle=-90]{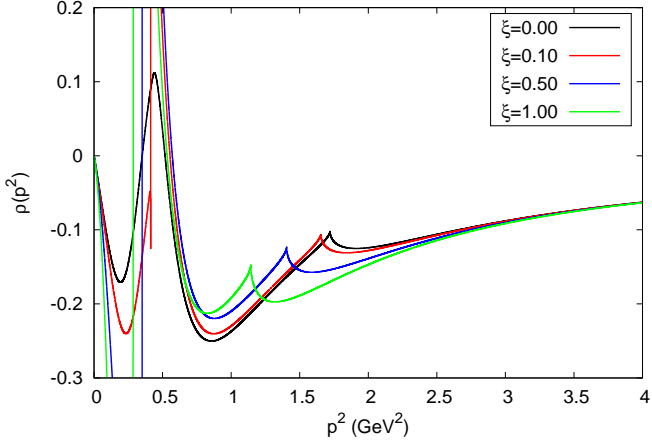}
\caption{The one-loop spectral density $\rho(p^2)$ is shown for different values of the gauge parameter.}
\end{figure}

\begin{figure}[t] \label{fig:spectdetail}
\centering
\includegraphics[width=0.35\textwidth,angle=-90]{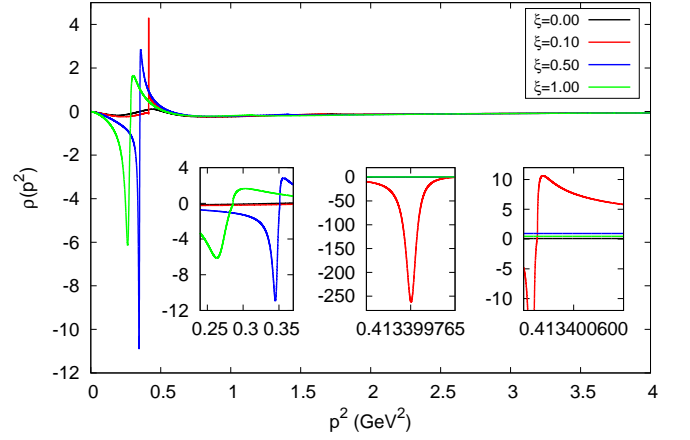}
\caption{The same curves of Fig.~11 on a different scale. The peaks of $\rho(p^2)$ are shown in the inserts on a very enlarged scale. 
The rightmost inserts both have a width of $2\times 10^{-7}$ GeV$^2$.}
\end{figure}

From a technical point of view, the new pole arises because of the logarithmic divergence of the real part of $F_\xi(-z^2/m^2)$ at the
branch point $x=m$ on the real axis. The divergence occurs because of the bad IR behavior of the crossed gluon loop, 
$\Pi_{2c}$ in Fig.~1, in the limit $p\to i m$, since the denominator in Eq.(\ref{Ipi2bx}) becomes
\BE
k^2\left[(k+p)^2+m^2\right]^{n+1}\to k^2\left[k^2+2k\cdot p\right]^{n+1}\sim k^{n+3}
\EE
if there are $n$ insertions of the counterterm in the transverse gluon line. Thus the integral diverges in the IR and the
divergence becomes worse and worse at higher orders, requiring some resummation which might cancel the divergence in the exact result.
For $n=1$ the divergence appears as a branch point at $s=-1$ for the logarithmic term $\log(1+s)$ of $F_\xi(s)$ in Eq.(\ref{Fxi}).
Near the branch point, for $x\approx m$ and any finite $\xi\not=0$, 
the real part of the inverse dressing function $\Psi(z)$, Eq.(\ref{Psi}), can be written as
\BE
\Rerm \Psi(z)\approx \Rerm \Psi_{reg}(z)+\xi A(m)\log\vert z-m\vert
\EE
where $\Psi_{reg}$ is the regular part and the prefactor $A(z)$ of the log is a rational function which is real on the real axis,
with $A(m)=-2/3$.
Then, taking $z=m+r e^{i\phi}$, the contour line $\Rerm \Psi=0$ is given by 
\BE
r\approx \exp \left(-\frac{\Rerm\Psi_{reg}(m)}{\xi A(m)} \right)=e^{-C/\xi}
\EE
which is a very small circle centered at $x=m$ on the real axis, with an exponentially small radius in the limit $\xi\to 0$ if
$C>0$. In the Feynman gauge, $\xi=1$, the contour line is just visible in Fig.~5  as a small black semi-circle centered at $x=m(1)=0.53$ GeV on
the real axis. It gets hardly visible for $\xi<0.5$. 

At the same branch point, the imaginary part of $F_\xi(s)$ has a large
discontinuous step yielding a change of the whole imaginary part 
\BE
\delta (\Im \Psi)\approx \xi \pi A(m)=-2.1\cdot \xi
\label{DIm}
\EE
which is quite larger than $\Im \Psi_{reg}(m)\approx 0.17$ and gives rise to a sharp change of sign at $x=m(\xi)$, even when $\xi$ is small, 
provided that $\xi>0.08$. 
On the complex plane, the discontinuous step is smeared out and the imaginary part $\Im \Psi$ changes sign on a contour line $\Im \Psi=0$ which 
originates from the branch point $x=m(\xi)$, just at the center of the circle $\Rerm \Psi=0$. The resulting contour line $\Im \Psi=0$ 
is visible in Fig.~5 as a solid red line ending at the center of the black semi-circle. The crossing point of the two contour lines
is the new pole that  appears for $\xi>0.08$. On the other hand, if $\xi<0.08$, the imaginary part $\Im \Psi$ changes sign below $x=m(\xi)$,  
out of the circle, the contour lines do not cross and the extra pole disappears when approaching the Landau gauge. 

By the previous analysis we conclude that the narrow peak of the spectral function must have a very small width, roughly 
given by the distance of the pole from the real axis $r\approx \exp(-C/\xi)$, getting smaller and smaller when $\xi\ll 1$, as shown in
Fig.~12. Moreover, $\Im \Psi$ and $\rho(p^2)$ change sign across the peak and the overall effect of the peak on the integral, in Eq.(\ref{KL}), 
is expected to be negligible.

It is likely that the sharp peak of the spectral function and the gauge dependent pole get smoothed
in the exact propagator since the Nielsen identity, Eq.(\ref{nielsen}) would forbid the existence of a pole which
depends on the gauge parameter $\xi$, unless the Green function $G^T(p)$ in Eq.(\ref{GT}) has a pole at the same point.
Having traced the source of the pole and found it related to the logarithmic divergence of the crossed graphs at $p^2=-m^2$,
we cannot exclude that the same divergence might occur in the ghost sector and in other Green functions.
Thus, in principle, we cannot rule out that the pole might be genuine, even if probably related to unphysical degrees of freedom of the
ghost sector.

\section{Discussion}

There is a growing consensus that QCD and Yang-Mills theory are self-contained theories that dynamically generate
their own infrared cutoff. The numerical simulations on the lattice have shown that the exact theory generates
a dynamical mass which screens the gluon interaction in the IR. Therefore, any continuum first-principle study should
reproduce the same results without the aid of any adjustable parameter, except for the overall energy scale that must come
from the phenomenology. It could be argued that, because of Gribov ambiguity, in $R_\xi$ gauge the Faddeev-Popov Lagrangian 
is just an approximation of the full theory. The approximation works very well in the usual perturbative approach but could
be out of control in the IR because of non-perturbative effects. A phenomenological parameter has been introduced by several authors
for locating the Gribov horizon, yielding an interaction-induced mass scale which screens 
the theory in the IR\cite{dudal08,capri,dudal11,tissier10,tissier11,tissier14,serreau,reinosa,pelaez}.
However, even averaging over Gribov copies, a dynamical mass is generated in the theory, as shown by the 
gauge-fixed lattice calculations in the Landau gauge. A recent analysis\cite{roberts17} has made clear that the dynamical mass would
be as effective as the Gribov parameter for screening the theory and that its dynamical appearance alone would eliminate the problem of
Gribov copies and complete the definition of the theory.

The same argument holds for the massive expansion which is a screened expansion from the beginning and can be safely used in the IR.
Having changed the expansion point, the gauge-fixed theory can be studied by plain perturbation theory and the agreement with the
lattice data shows that, when the expansion is optimized, higher order graphs are very small and negligible. Thus, ignoring the
Gribov ambiguity does not seem to be a problem as far as perturbation theory works well. Again, it is a consequence of the
dynamical mass that screens the theory, yielding a self-contained perturbative description from first principles.

It is not surprising that, without using any adjustable parameter and without modifying the original gauge-fixed Lagrangian, 
the massive expansion predicts the same pole structure which was found by the refined Gribov-Zwanziger model\cite{dudal08,capri,dudal11}. 
The two approaches are very different but they study the same identical physical system, so that if both are valid approximations they must 
reach the same conclusions. Moreover, our analysis supports the physical relevance of the principal part:
having established its gauge parameter independence\cite{dudalR}, we argue that the simple rational part $\Delta^P(p)$ might play an
important role in the phenomenology, more than the (small) gauge dependent spectral density.
The conclusions of the present work would be enforced by a comparison with position-space lattice data, because of their
sensitivity to the analytical structure of the propagator. Unfortunately, at the moment, for a generic covariant gauge, 
no such data are available.

An apparent drawback of the massive expansion is that the BRST invariant action is arbitrarily splitted in two parts that 
are not BRST invariant. The Nielsen identities cannot be satisfied exactly at any finite order of the expansion.
However, because of the spurious dependence of the approximation on the subtraction point $\mu/ m$, the expansion can be
optimized by enforcing the gauge parameter invariance of the pole structure. Thus, the extension to $R_\xi$ gauge, not only
gives new information on the gluon propagator in a generic gauge, but also provides a unique way to fix the optimal expansion even
in the Landau gauge. 
The good agreement with the available lattice data, which is reached without any fit of adjustable parameters, 
increases our confidence in the general validity of the method as a first-principle benchmark for more phenomenological models.

\acknowledgments

We are in debt to David Dudal for suggesting the proof of gauge parameter independence of the residues.
We also thank Orlando Oliveira for sharing with us the lattice data of Ref.\cite{duarte}.

\appendix
\section{one-loop graphs}

In this appendix, explicit analytical expressions are derived for the one-loop polarization graphs of Fig.1.
The graphs are evaluated using the free-particle (gauge-dependent) propagator of Eq.(\ref{Deltam}) and inserting
the transverse counterterm of Eq.(\ref{dG2}) as a new two-point vertex which is shown as a cross in the figure.
We refer to the graphs with one insertion of the counterterm as crossed graphs.

\subsection{Graphs $\Pi_{1b}$, $\Pi_{1c}$ and $\Pi_{1d}$ (tadpoles)}

In the Euclidean space, the constant tadpole $\Pi_{1b}$ can be written as
\BE
\Pi_{1b}=-\frac{N g^2 (d-1)^2}{d} \int\kkd\frac{1}{k^2+m^2}
\label{Ipi1b}
\EE
having dropped the longitudinal loop which is scaleless and vanishes in dimensional regularization.
Setting  $d=4-\epsilon$, in the $\overline{MS}$ scheme, 
\BE
\Pi_{1b}=\frac{3}{4} \frac{(3 Ng^2)}{(4 \pi)^2}\>m^2\>\left(\frac{2}{\epsilon}+\log\frac{\mu^2}{m^2}+C\right)
\label{Pi1b}
\EE
where $C$ is a constant which depends on the regularization scheme. 

The crossed graphs do not contain any longitudinal gluon line since the counterterm $\delta \Gamma$ is transverse
in Eq.(\ref{dG2}).
The graph $\Pi_{1c}$ can be written as a derivative 
\BE
\Pi_{1c}=-m^2\frac{{\partial}\Pi_{1b}}{{\partial} m^2}=
-\frac{3}{4}\frac{(3 Ng^2)}{(4 \pi)^2} \>m^2\>\left(\frac{2}{\epsilon}+\log\frac{\mu^2}{m^2}+C-1\right).
\EE

As expected, the diverging terms cancel in the sum $\Pi_{1b}+\Pi_{1c}$. 
The double-crossed tadpole $\Pi_{1d}$ is finite and including its symmetry factor it reads
\BE
\Pi_{1d}= \frac{1}{2} m^4\frac{{\partial^2}\Pi_{1b}}{{\partial} (m^2)^2}=   -\frac{3}{8} \,\frac{(3 Ng^2)}{(4 \pi)^2} \>m^2
\EE
so that the sum of the constant graphs is
\BE
\Pi_{1b}+\Pi_{1c}+\Pi_{1d}=\frac{3}{8} \,\frac{(3 Ng^2)}{(4 \pi)^2} \>m^2.
\label{CG}
\EE 

\subsection{Ghost loop $\Pi_{2a}$}

The ghost loop $\Pi_{2a}$ is a standard graph and does not depend on $\xi$. In the Euclidean space it 
is given by the 
integral\cite{genself}
\BE
\Pi_{2a}(p)=-\frac{Ng^2}{(d-1)}
\int\kkd\>\frac{k_\perp^2}{k^2(p+k)^2}.
\label{Ipi2a}
\EE
The integral is straightforward and setting $d=4-\epsilon$ the diverging part is
\BE
\Pi^\epsilon_{2a} (p)=\frac{(3 Ng^2)}{(4 \pi)^2}\,\frac{ p^2}{36}
\left(\frac{2}{\epsilon}+\log \frac{\mu^2}{m^2}\right)
\label{2aeps}
\EE
while the finite part reads
\BE
\Pi^f_{2a} (p)=\frac{(3 Ng^2)}{(4 \pi)^2}\,\frac{ m^2}{36}
\left(C_0 s-s\log s\right)
\label{2af}
\EE
where $s=p^2/m^2$ and the constant $C_0$ depends on the regularization scheme.

\subsection{Gluon loop $\Pi_{2b}$}

The gluon loop $\Pi_{2b}$ can be written as 
\BE
\Pi_{2b}(p)=\Pi_{2b}^0(p)+\xi \,\Pi_{2b}^\xi(p)+\xi^2 \,\Pi_{2b}^{\xi\xi}(p)
\EE
where $\Pi_{2b}^0(p)$ is the graph in the Landau gauge, $\xi=0$. 
In the Euclidean space, setting $d=4$, it reads\cite{genself}
\BE
\Pi^0_{2b} (p)=\frac{Ng^2}{6}\int\kkk\frac{k_\perp^2{\cal F}^0(k,p)}{(k^2+m^2)[(k+p)^2+m^2]}
\label{Ipi2b}
\EE
where $k_\perp^2=[k^2-(k\cdot p)^2/p^2]$ and the kernel ${\cal F}^0$ can be derived by the explicit
expressions of Ref.\cite{genself}
\BE
{\cal F}^0 (k,p)=\frac{10(k^2+p^2)+(k+p)^2}{k^2}+\frac{p^4+10p^2k^2+k^4}{k^2(k+p)^2}.
\label{kernF0}
\EE

It is useful to decompose it as
\BE
\frac{{\cal F}^0 (k,p)}{12}=\frac{k^2+p^2}{k^2}+\frac{p^2}{(k+p)^2}-\frac{p^2k_\perp^2}{3(k+p)^2 k^2}
\label{kernF0d}
\EE
and using the identity
\BE
\frac{1}{q^2 (q^2+m^2)}=\frac{1}{m^2}\left[\frac{1}{q^2}-\frac{1}{q^2+m^2}\right]
\EE
the graph can be split as 
\BE
\Pi^0_{2b}(p)=2Ng^2\left[I_A(p)+2I_B(p)+I_C(p)\right]
\EE
where
\begin{align}
I_A(p)&=\int\kkk\frac{k_\perp^2\left(\displaystyle{1-\frac{2p^2}{m^2}-\frac{p^2k_\perp^2}{3m^4}}\right)}{(k^2+m^2)[(k+p)^2+m^2]}\nn
\end{align}
\begin{align}
I_B(p)&=\frac{p^2}{m^2}\int\kkk\frac{k_\perp^2\left(\displaystyle{1+\frac{k_\perp^2}{3m^2}}\right)}{k^2[(k+p)^2+m^2]}\nn
\end{align}
\begin{align}
I_C(p)&=-\frac{p^2}{3m^4}\int\kkk\frac{k_\perp^4}{k^2(k+p)^2}.
\end{align}

The integrals can be evaluated analytically\cite{ptqcd2,tissier10,tissier11} by dimensional regularization 
for $d=4-\epsilon$, yielding
a diverging part
\BE
{\Pi^0}^\epsilon_{2b}(p)=-\frac{3 Ng^2}{(4 \pi)^2} \left( m^2-\frac{25}{36} p^2\right)
\left(\frac{2}{\epsilon}+\log\frac{\mu^2}{m^2}\right)
\label{2beps}
\EE
and a finite part
\begin{widetext}
\BE
{\Pi^0}^f_{2b}=\frac{3 Ng^2}{(4 \pi)^2}\,\frac{ m^2}{72}\left[\frac{2}{s}+C_1+C_2 s+s^3\log s -s L_A(s)-s L_B(s)\right]
\label{2bf}
\EE
where $C_1$, $C_2$ are constants which depend on the regularization scheme,
$s=p^2/m^2$ and $L_A$, $L_B$ are the logarithmic functions
\begin{align}
L_A(s)&=(s^2-20s+12)\left(\frac{4+s}{s}\right)^{3/2}
\log\left(\frac{\sqrt{4+s}-\sqrt{s}}{\sqrt{4+s}+\sqrt{s}}\right)\nn\\
L_B(s)&=\frac{2(1+s)^3}{s^3}(s^2-10s+1)\log(1+s).
\label{logs}
\end{align}
\end{widetext}

The other terms, $\Pi_{2b}^\xi$ and  $\Pi_{2b}^{\xi\xi}$, arise by substituting one and two
transverse lines, respectively, with the longitudinal ones. By the general scheme of Ref.\cite{genself},
for $d=4$, they follow as
\begin{align}
\Pi^\xi_{2b} (p)&=\frac{Ng^2}{6}\int\kkk\frac{{\cal F}^{0\xi}(k,p)}{(k^2+m^2)(k+p)^2}\nn\\
&\quad+\frac{Ng^2}{6}\int\kkk\frac{{\cal F}^{\xi0}(k,p)}{k^2\left[(k+p)^2+m^2\right]}
\label{Ipi2bx}
\end{align}
\begin{align}
\Pi^{\xi\xi}_{2b} (p)&=\frac{Ng^2}{6}\int\kkk\frac{{\cal F}^{\xi\xi}(k,p)}{k^2(k+p)^2}\nn\\
\label{Ipi2bxx}
\end{align}
where
\begin{align}
{\cal F}^{0\xi}(k,p)&=\frac{(3k^2-k_\perp^2)(k^2-p^2)^2}{k^2(k+p)^2}=\nn\\
&=3(k+p)^2-\frac{(10p^2+k^2)k_\perp^2}{(k+p)^2}\nn\\
&\qquad -\frac{p^4k_\perp^2}{k^2(k+p)^2}-12(p\cdot k),\nn\\
{\cal F}^{\xi0}(k,p)&=3k^2+12p^2+12(k\cdot p)-k_\perp^2\nn\\
&\qquad\qquad -k_\perp^2\left[\frac{11p^2 +2(k\cdot p)}{k^2}
+\frac{p^4}{(k+p)^2k^2}\right],\nn\\
{\cal F}^{\xi\xi}(k,p)&=\frac{p^4k_\perp^2}{k^2(k+p)^2}.
\label{kernels}
\end{align}

The quadratic term is trivial since the integral $\Pi^{\xi\xi}_{2b}$ is scaleless and by a dimensional argument
$\Pi^{\xi\xi}_{2b}(p)={\rm const} \times p^2$. The constant can be absorbed by a finite wave function renormalization and
the term can be ignored.

The two integrals in Eq.(\ref{Ipi2bx}) must be the same, as can be easily seen by substituting $k\to (-k-p)$ 
in Eq.(\ref{kernels}). Taking twice the explicit expression of ${\cal F}^{\xi0}$, the integral can be written as 
\BE
\Pi^{\xi}_{2b}(p)=Ng^2\left[I^\xi_A(p)+I^\xi_B(p)+I^\xi_C(p)+I^\xi_D(p)\right]
\EE
where
\begin{align}
I^\xi_A(p)&=\int\kkk\frac{1}{[(k+p)^2+m^2]}\nn\\
I^\xi_B(p)&=\frac{1}{3}\int\kkk\frac{12p^2+12(k\cdot p)-k_\perp^2}{k^2[(k+p)^2+m^2]}\nn\\
I^\xi_C(p)&=-\frac{p^4}{3m^2}\int\kkk\frac{k_\perp^2}{(k^2)^2(k+p)^2}\nn\\
I^\xi_D(p)&=\frac{1}{3}\int\kkk\frac{k_\perp^2\left[\displaystyle{\frac{p^4}{m^2}}-11p^2-2(k\cdot p)\right]}{(k^2)^2[(k+p)^2+m^2]}
\end{align}

By dimensional regularization, taking $d=4-\epsilon$, the integrals can be evaluated analytically in the $\overline{MS}$ scheme.
The first integral is the same occurring in Eq.(\ref{Pi1b}) 
\BE
I^\xi_A(p)=-\frac{m^2}{(4 \pi)^2}\>\left(\frac{2}{\epsilon}+\log\frac{\mu^2}{m^2}+C_A\right)
\label{IxiA}
\EE
The other integrals are
\begin{align}
&I^\xi_B(p)=\frac{2m^2}{(4 \pi)^2}\>\bigg\{\left(\frac{2}{\epsilon}+\log\frac{\mu^2}{m^2}\right)\left(\frac{25}{24}s+\frac{1}{8}\right)
+C_B s+ C_B^\prime\nn\\
&\quad+\frac{24 s (1-s^2)-(1+s)^3}{24 s^2}\, \log(1+s)+\frac{1}{24 s}\bigg\}
\label{IxiB}
\end{align}
\begin{align}
&I^\xi_C(p)=-\frac{m^2 s^2}{4(4 \pi)^2}\>\bigg\{
\left(\frac{2}{\epsilon}+\log\frac{\mu^2}{m^2}+C_C\right) - \log s  \bigg\}
\label{IxiC}
\end{align}
\begin{align}
&I^\xi_D(p)=\frac{m^2 s^2}{4(4 \pi)^2}\bigg\{\left(\frac{2}{\epsilon}+\log\frac{\mu^2}{m^2}+C_C\right)\nn\\
&\qquad\qquad\qquad+\frac{(1-s^2)}{s^2} \log(1+s)\bigg\}\nn\\
&\quad+\frac{m^2}{12(4 \pi)^2}\> \bigg\{
-31s\left(\frac{2}{\epsilon}+\log\frac{\mu^2}{m^2}+C_D\right) + C_D^\prime\nn\\
&\quad\quad+\frac{(1+s)(31s^2-31s+4)}{s^2} \log(1+s)-\frac{4}{s} \bigg\}
\label{IxiD}
\end{align}
where all constants $C_X$, $C_X^\prime$ depend on the regularization scheme. 
In Eq.(\ref{IxiD}), the first two lines arise from the $p^4$ term of $I^\xi_D(p)$ and
the diverging term cancels the corresponding divergence of $I^\xi_C(p)$ in Eq.(\ref{IxiC}).

Adding up the different integrals we obtain a diverging part
\BE
{\Pi^\xi}^\epsilon_{2b}(p)=-\frac{Ng^2}{4(4 \pi)^2} \left( 3m^2+2 p^2\right)
\left(\frac{2}{\epsilon}+\log\frac{\mu^2}{m^2}\right)
\label{2bxieps}
\EE
and a finite part
\BE
{\Pi^\xi}^f_{2b}=\frac{Ng^2}{(4 \pi)^2}\,\frac{ m^2}{4}\left[
\frac{(1+s)(1-s)^3}{s^2}\log(1+s)+s^2\log s- \frac{1}{s}\right]
\label{2bxif}
\EE
where we have omitted the irrelevant constants.

Finally, the gluon loop has the following structure
\BE
\Pi_{2b}=\left[{\Pi^0}^\epsilon_{2b}+\xi{\Pi^\xi}^\epsilon_{2b}\right]+\left[{\Pi^0}^f_{2b}+\xi{\Pi^\xi}^f_{2b}\right].
\EE

\subsection{Standard one-loop graphs}

The standard one-loop result of perturbation theory does not contain any contribution from the crossed graphs.
In a generic linear covariant gauge, the standard one-loop polarization $\Pi_1(p)$ is obtained as the sum
\BE
\Pi_1(p)=\Pi_{1b}+\Pi_{2a}(p)+\Pi_{2b}(p)
\EE
and summing up the explicit expressions reported above, we find a diverging part
\BE
\Pi^\epsilon_1(p)
=\frac{Ng^2}{(4 \pi)^2}\left(\frac{2}{\epsilon}+\log\frac{\mu^2}{m^2}\right)
\left[p^2\left(\frac{13}{6}-\frac{\xi}{2}\right)-\frac{3}{4} m^2 \left(1+\xi\right)\right]
\label{Pi1eps}
\EE
and a finite part
\BE
\Pi^f_1(p)=-\frac{Ng^2}{4!\,(4 \pi)^2}\, p^2\,\left\{ C_p+\frac{1}{s}\bigg[
C_m+f(s)+\xi f_\xi(s)\bigg]\right\}
\label{Pi1f}
\EE
where 
\begin{align}
f(s)&=s\left[L_A(s)+L_B(s)+(2-s^2)\log s-2 s^{-2}\right]\nn\\
f_\xi(s)&=6\left[s^{-1}-s^2\log s-\frac{(1+s)(1-s)^3}{s^2} \log(1+s)\right].
\label{f}
\end{align}
In the limit $m\to 0$ the diverging part in Eq.(\ref{Pi1eps}) agrees with the well known result of perturbation
theory\cite{peskin}. In the limit $\xi\to 0$ the finite part in Eq.(\ref{Pi1f}) gives the known result in the
Landau gauge\cite{tissier10,tissier11}.
The constants $C_m$ and $C_p$ are arbitrary since they depend on the regularization scheme
and on the arbitrary energy scale $\mu$ in Eq.(\ref{Pi1eps}). In the standard perturbation theory,
they are the finite parts resulting from the cancellation of the divergences by mass and wave function
renormalization, respectively. In pure Yang-Mills theory, there is no mass term in the original Lagrangian and
no mass renormalization for the cancellation. However, all constant mass terms cancel exactly by inclusion of the
crossed graphs.

\subsection{Total polarization (including the crossed graphs)}

All crossed graphs, containing one insertion of the transverse mass counterterm, can be added to the total one-loop
polarization by a simple derivative, as discussed above, for the tadpole.
The sum of all graphs in Fig.~1 follows as
\BE
\Pi_{tot}(p)=\left(1-m^2\frac{\partial}{\partial m^2}\right)\Pi_1(p)+\Pi_{1d}.
\EE
Using the identity
\BE
\left(1-m^2\frac{\partial}{\partial m^2}\right)=\left(1+s\frac{\partial}{\partial s}\right)
\EE
and adding up the terms, we obtain a total diverging part
\BE
\Pi^\epsilon_{tot}(p)
=\frac{Ng^2}{(4 \pi)^2}\left(\frac{2}{\epsilon}+\log\frac{\mu^2}{m^2}\right)
p^2\left(\frac{13}{6}-\frac{\xi}{2}\right)
\label{Piteps}
\EE
and a total finite part
\begin{align}
\Pi^f_{tot}(p)&=-3\frac{Ng^2}{(4 \pi)^2} 
\, p^2\,\Bigg\{\frac{1}{s}\left(\frac{5}{8}+\frac{\xi}{4}\right)\nn\\
&\qquad+\frac{1}{3\cdot 4!}\left[
f^\prime (s)+\xi f^\prime_\xi(s)\right]+{\rm const.}\Bigg\}
\label{Pitf}
\end{align}
where $f^\prime (s)$ and $f_\xi^\prime (s)$ are the derivatives of the functions $f (s)$ and $f_\xi (s)$, respectively.

Finally, inserting the polarization in Eq.(\ref{dressD}) and canceling the divergence by the usual wave function renormalization,
the renormalized dressed propagator reads
\BE
\Delta(p)=\frac{Z}{p^2\left[F(s)+\xi F_\xi (s) + F_0\right]}
\label{propR}
\EE
where $Z$ is an arbitrary  finite renormalization factor, $F_0$ is a finite additive constant and the adimensional functions 
$F$, $F_\xi$ do not depend on any parameter and are defined as
\begin{align}
F(s)&=\frac{5}{8s}+\frac{1}{3\cdot 4!}\,f^\prime (s)\nn\\
F_\xi(s)&=\frac{1}{4s}+\frac{1}{3\cdot 4!}\,f_\xi^\prime (s).
\end{align}
Their explicit expressions follow by the simple derivative
of Eq.(\ref{f}). The function $F(x)$ was first derived in Refs.\cite{ptqcd,ptqcd2} and it reads
\begin{align}
F(x)&=\frac{5}{8x}+\frac{1}{72}\left[L_a+L_b+L_c+R_a+R_b+R_c\right]\nn\\
\label{Fx}
\end{align}
where the logarithmic functions $L_x$ are
\begin{align}
L_a(x)&=\frac{3x^3-34x^2-28x-24}{x}\>\times\nn\\
&\times\sqrt{\frac{4+x}{x}}
\log\left(\frac{\sqrt{4+x}-\sqrt{x}}{\sqrt{4+x}+\sqrt{x}}\right)\nn\\
L_b(x)&=\frac{2(1+x)^2}{x^3}(3x^3-20x^2+11x-2)\log(1+x)\nn\\
L_c(x)&=(2-3x^2)\log(x)
\label{logsA}
\end{align}
and the rational parts $R_x$ are
\begin{align}
R_a(x)&=-\frac{4+x}{x}(x^2-20x+12)\nn\\
R_b(x)&=\frac{2(1+x)^2}{x^2}(x^2-10x+1)\nn\\
R_c(x)&=\frac{2}{x^2}+2-x^2.
\label{rational}
\end{align}

The explicit expression of $F_\xi(x)$ is
\begin{align}
F_\xi(x)&=\frac{1}{4x}-\frac{1}{12}\bigg[
2x\log x-\frac{2(1-x)(1-x^3)}{x^3}\log(1+x)\nn\\
&\quad +\frac{3x^2-3x+2}{x^2}\bigg]
\label{Fxi}
\end{align}
and has the leading behavior in the limit $x\to 0$
\BE
F_\xi(x)=\frac{1}{4x}-\frac{1}{9}-\frac{x}{6}\log x+{\cal O}(x).
\EE
In the same IR limit, the transverse propagator is finite 
\BE
\Delta (0)=\frac{Z}{M_\xi^2}
\EE
and the mass parameter $M_\xi^2$ is defined as
\BE
M_\xi^2=\frac{5m^2}{8}\left(1+\frac{2}{5}\,\xi\right).
\EE

In the limit $x\to\infty$, the asymptotic UV behavior is
\begin{align}
F_\xi(x)&\sim -\frac{1}{6}\log x\nn\\
F(x)&\sim \frac{13}{18}\log x
\end{align}
and by Eqs.(\ref{Pitf}),(\ref{propR}), the standard one-loop behavior is recovered in the UV for the total polarization
and the dressed propagator  
\begin{align}
\Pi^f_{tot}(p)&\sim -\frac{Ng^2}{(4 \pi)^2} \, p^2\,\Bigg(\frac{13}{6}-\frac{\xi}{2}\Bigg) \log\frac{p^2}{\mu^2}\nn\\
\frac{Z}{\Delta (p)}&\sim \, p^2\, \Bigg(\frac{13}{6}-\frac{\xi}{2}\Bigg) \log\frac{p^2}{\mu^2}.
\label{UV}
\end{align}

The discussion on gauge invariance requires the derivatives of the functions $F(x)$ and $F_\xi(x)$.
The derivative of $F(x)$ reads
\BE
F^\prime (x)=-\frac{5}{8x^2}+\frac{1}{72}\left[L^\prime_a+L^\prime_b+L^\prime_c+R(x)\right]
\label{der1}
\EE
where the logarithmic functions $L^\prime_x$, for $x=a,b,c$, are
\begin{align}
L^\prime_a(x)&=\frac{6x^4-16x^3-68x^2+80x+144}{x^2(x+4)}\>\times\nn\\
&\times\sqrt{\frac{4+x}{x}}
\log\left(\frac{\sqrt{4+x}-\sqrt{x}}{\sqrt{4+x}+\sqrt{x}}\right)\nn\\
L^\prime_b(x)&=\frac{4(1+x)}{x^4}(3x^4-10x^3+10x^2-10x+3)\log(1+x)\nn\\
L^\prime_c(x)&=-6x \log x
\end{align}
and $R(x)$ is the sum of all the rational terms coming out from the derivatives
\BE
R(x)=\frac{12}{x}+\frac{106}{x^2}-\frac{12}{x^3}.
\EE
The derivative of $F_\xi(x)$ reads
\begin{align}
F^\prime_\xi(x)&=\frac{x^4+2x-3}{6x^4}\log(1+x)
-\frac{1}{6}\log x\nn\\
&\quad +\frac{(1-x)(1-x^3)}{6x^3(1+x)}
+\frac{1}{3x^3}-\frac{1}{2x^2}-\frac{1}{6}.
\label{der2}
\end{align}

\end{document}